\documentstyle[aps,epsfig]{revtex}

\draft

\begin{document}

\twocolumn[
\hsize\textwidth\columnwidth\hsize\csname@twocolumnfalse\endcsname

\title{X-ray Investigation of the Magneto-elastic Instability of 
$\alpha ^{\prime }$-NaV$_{2}$O$_{5}$}
\author{S. Ravy}
\address{Laboratoire de physique des solides, CNRS URA 02, B\^{a}t. 510, 
Universit\'{e}
Paris-sud, 91405 Orsay Cedex, France}
\author{J. Jegoudez and A. Revcolevschi}
\address{Laboratoire de chimie des solides, CNRS URA 446, B\^{a}t. 414, 
Universit\'{e} Paris-sud,
91405 Orsay Cedex, France}
\maketitle

\begin{abstract}
We present an X-ray diffuse scattering study of the pretransitional
structural fluctuations of the magneto-elastic transition in
 $\alpha ^{\prime }$-NaV$_{2}$O$_{5}$. 
This transition is characterized by the appearance
below $T_{sp}\sim 35K$ of satellite reflections at the reduced 
wave vector (1/2,1/2,1/4). A large regime of structural fluctuations is 
measured up to 90 K. These fluctuations are three dimensional between 
$T_{sp}$ and $\sim 50K$ and quasi-one dimensional above $\sim 60K$.
At 40 K the anisotropy ratio ( $\xi _{b} : \xi _{a} : \xi _{c}$) is
found to be ( 3.8 : 1.8 : 1 ), which reveals the 
importance of transverse interactions in the stabilization of the low 
temperature phase. We discuss our results within the framework of recent 
theories dealing with the simultaneous occurrence of a charge ordering, 
a spin gap and a lattice distortion in this intriguing compound.
\end{abstract}

\vfill
\twocolumn
\vskip.5pc ]

\narrowtext

First observed in the 70's in the organic compounds MEM(TCNQ)$_{2}$\cite{Hui}
and TTF-CuBDT\cite{Bray},\cite{Monc} and later in the charge transfer salts
(TMTTF)$_{2}$PF$_{6}$\cite{Creuzet}, (TMTTF)$_{2}$AsF$_{6}$\cite{Laver} and
(BCP-TTF)$_{2}$X\cite{Ducasse},\cite{Liu} (where X=PF$_{6}$ and AsF$_{6}$),
the spin-Peierls (SP) transition regained interest after its discovery in
the inorganic system CuGeO$_{3}$\cite{Hase}. The SP ground state is a
magneto-elastic distortion of a one-dimensional (1D) antiferromagnetic chain
of spin-1/2. Below the transition temperature $T_{c}$ a singlet-triplet gap
opens in the magnetic excitation spectrum, accompanied by a lattice doubling
of the chain. Theoretically, the SP transition is driven by one-dimensional
antiferromagnetic fluctuations coupled to the lattice via the spin-phonon
coupling. Accordingly, a strong and anisotropic regime of lattice
fluctuations is expected to occur. Nevertheless, SP systems do not all
exhibit such a regime of fluctuations as we already emphasized in a previous
study\cite{Liu2}. In MEM(TCNQ)$_{2}$ and TTF-CuBDT the temperature
dependence of the magnetic susceptibility $\chi (T)$ down to $T_{c}$ is well
accounted for by a Bonner-Fisher (BF) law of a one-dimensional S=1/2
Heisenberg chain. However, an isotropic x-ray diffuse scattering at the
location of the low temperature superlattice reflection remains until high
temperature\cite{Liu2}\cite{Monc}. On the other hand, in (BCP-TTF)$_{2}$AsF$%
_{6}$ a quasi-one dimensional fluctuation regime is observed and $\chi (T)
$ exhibits a sizeable deviation from the BF law, interpreted as due to the
influence of the lattice fluctuations on the spin degrees of freedom\cite
{Dumoulin}. As far as CuGeO$_{3}$ is concerned, an anisotropic fluctuation
regime has been observed until $\sim 2.5T_{c}$\cite{Schoeffel}. Here
again, $\chi (T)$ does not follow the BF law due to either deviation from
one-dimensionality or to the presence of sizeable next-nearest neighbor
interactions. All these results show that, although the spin-Peierls 
transition concept was successfully used to describe all these
magneto-elastic transitions, the experimental situation is much more
intricate.

The inorganic compound $\alpha ^{\prime }$-NaV$_{2}$O$_{5}$ has recently
been shown to undergo a magneto-elastic phase transition at $T_{c}=34K$\cite
{Isobe}. Above $T_{c}$, $\chi (T)$ follows remarkably the BF law (with the
exchange constant $J/k_{B}=$280 K) and the abrupt decrease of $\chi (T)$
below $T_{c}$ is accompanied by a lattice doubling observed by x-ray
diffraction\cite{Fujii}. Moreover, the existence of a spin gap $\Delta $ of $%
\sim 10$ $meV$ was evidenced by inelastic neutron scattering carried out on 
a powder sample\cite{Fujii}. These observations suggested that 
$\alpha ^{\prime }$-NaV$_{2}$%
O$_{5}$ might undergo an ''ideal'' SP transition, suitable to test more
precisely the current theory of 1D quantum spin systems. The orthorhombic
structure (a=11.311 \AA , b=3.61 \AA\ and c=4.80 \AA ) consists of
double-rows of edge-sharing VO$_{5}$ pyramids running in the ${\bf b}$
direction, and oriented alternately upward and downward in the ${\bf a}${\bf %
\ }direction. These (${\bf a}${\bf ,}${\bf b}$) layers are separated by Na$%
^{+}$ cations in the ${\bf c}${\bf \ }direction. The first description of
the structure\cite{Carpy} used the non-centrosymmetric space group $P2_{1}mn$%
, which described the compound as an assembly of alternating magnetic V$^{4+}$
(S=1/2) and non-magnetic V$^{5+}$ (S=0) chains. However, careful
reinvestigations of the structure\cite{Centro},\cite{Smolinski}, have shown
that the actual space group is the centrosymmetric one, $Pmmn$, which implies
that all the V-chains are equivalent. The same conclusion was drawn from $%
^{51}$V NMR measurements\cite{Omaha}. Only one set of V sites is found in
the spectrum above T$_{c}$, while two inequivalent sets of V sites, assigned
to V$^{4+}$ and V$^{5+}$, are observed below the transition, evidencing a
charge ordering process associated with the phase transition. These
observations have made more difficult the description of the compound as made 
of an assembly of spin-1/2 chains. Furthermore, in contrast with other SP 
systems where the spin gap has been measured, the ratio $2\Delta /kT_{c}\ $
is equal to $6.5$ instead of the BCS-value $3.52$. Additional thermodynamic
measurements\cite{Köppen} have also indicated the occurrence of two phase
transitions very close to $T_{c}$, one of them being first-order.

All these results show that $\alpha ^{\prime }$-NaV$_{2}$O$_{5}$ undergoes a
magneto-elastic transition far to be correctly described by the conventional
spin-Peierls scenario. Especially interesting is the concomitant occurrence
(in the experimental limits given in ref. \cite{Köppen}) of a charge
ordering, a spin gap and a lattice doubling. Theoretical studies have
recently addressed this issue\cite{Smolinski}\cite{Seo}\cite{Nishimoto}\cite
{Mostovoy}\cite{Thalmeier}. An attractive model considers the system as a
quarter-filled ladder compound\cite{Seo}\cite{Smolinski}\cite{Nishimoto}\cite
{Mostovoy}. By taking into account the on-site and intersite Coulomb
interactions, different types of charge ordering are considered. In the
'chain' model, the V$^{4+}$ are ordered along one leg of the ladder but a
conventional spin-Peierls mechanism is still invoked to stabilize the
lattice distortion leading to the spin gap\cite{Nishimoto}\cite{Thalmeier}.
In a second model in which the V$^{4+}$ form a 'zig-zag' pattern along the
ladder, the spin gap is seen as a natural consequence of the transverse
order, although the exact mechanism of the spin pairing is controversial\cite
{Seo}\cite{Mostovoy}. As far as the lattice coupling is concerned, Mostovoy 
{\it et al}\cite{Mostovoy} emphasized its importance in stabilizing the 
'zig-zag' ordering with respect to the 'chain' one. Besides, the role of the
transverse interactions has not been investigated in detail in these
studies. In order to clarify these points, we have performed an x-ray study
of the pretransitional structural fluctuations of $\alpha ^{\prime }$-NaV$%
_{2}$O$_{5}$.

\begin{figure}
\centerline{\epsfig{file=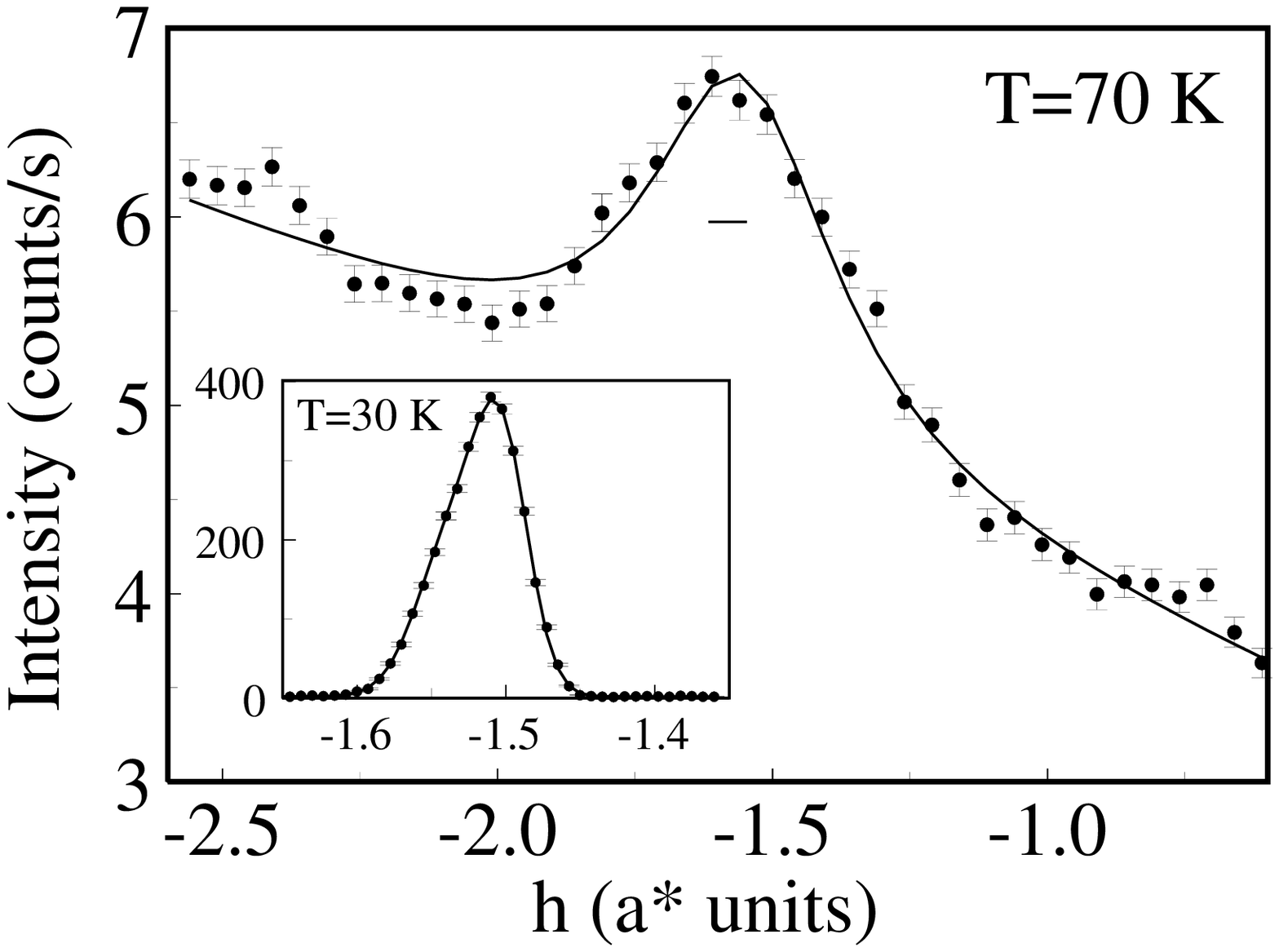,angle=0.0,height=5cm,width=7cm}}
\centerline{\epsfig{file=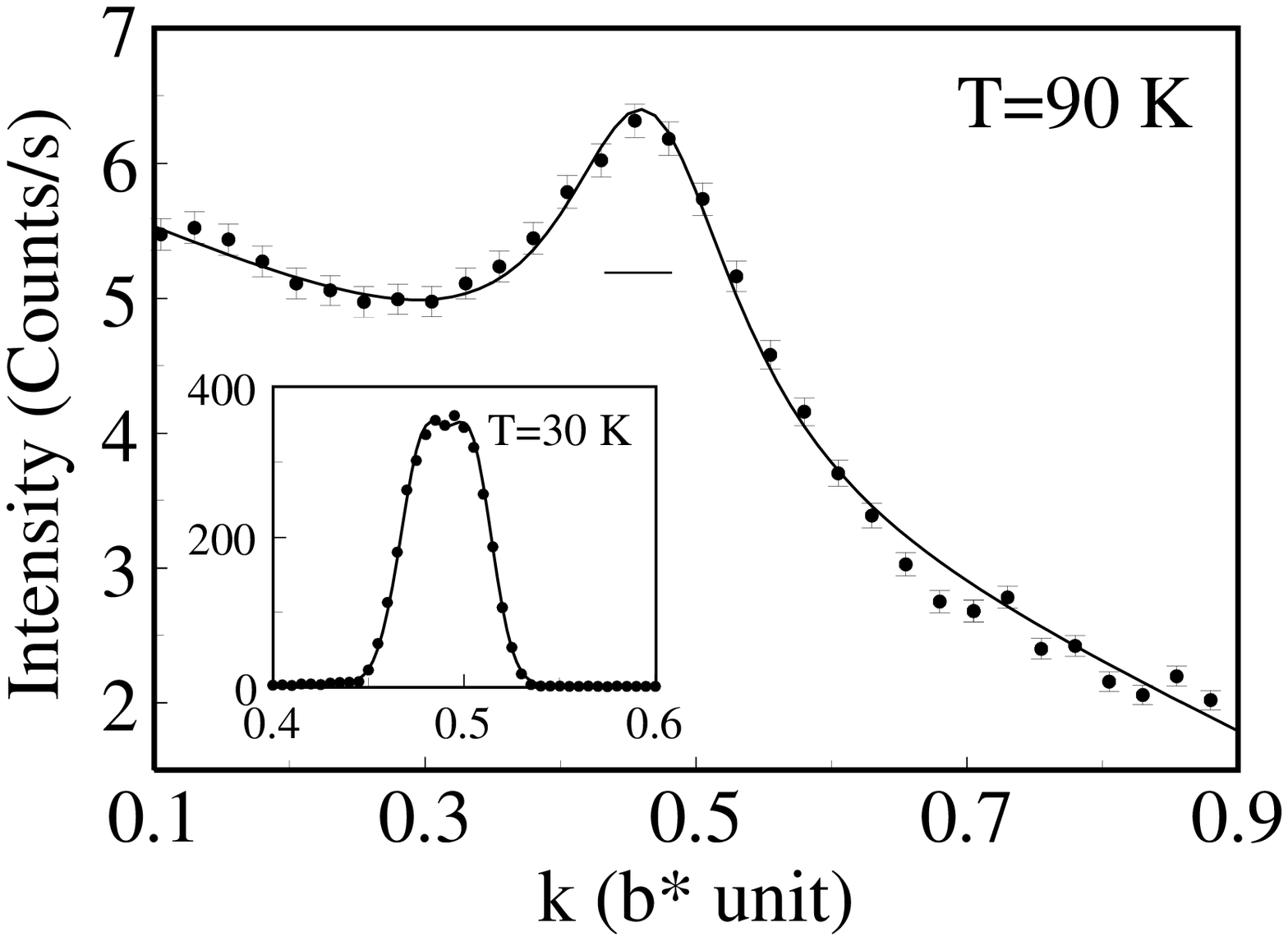,angle=0.0,height=5cm,width=7cm}}
\centerline{\epsfig{file=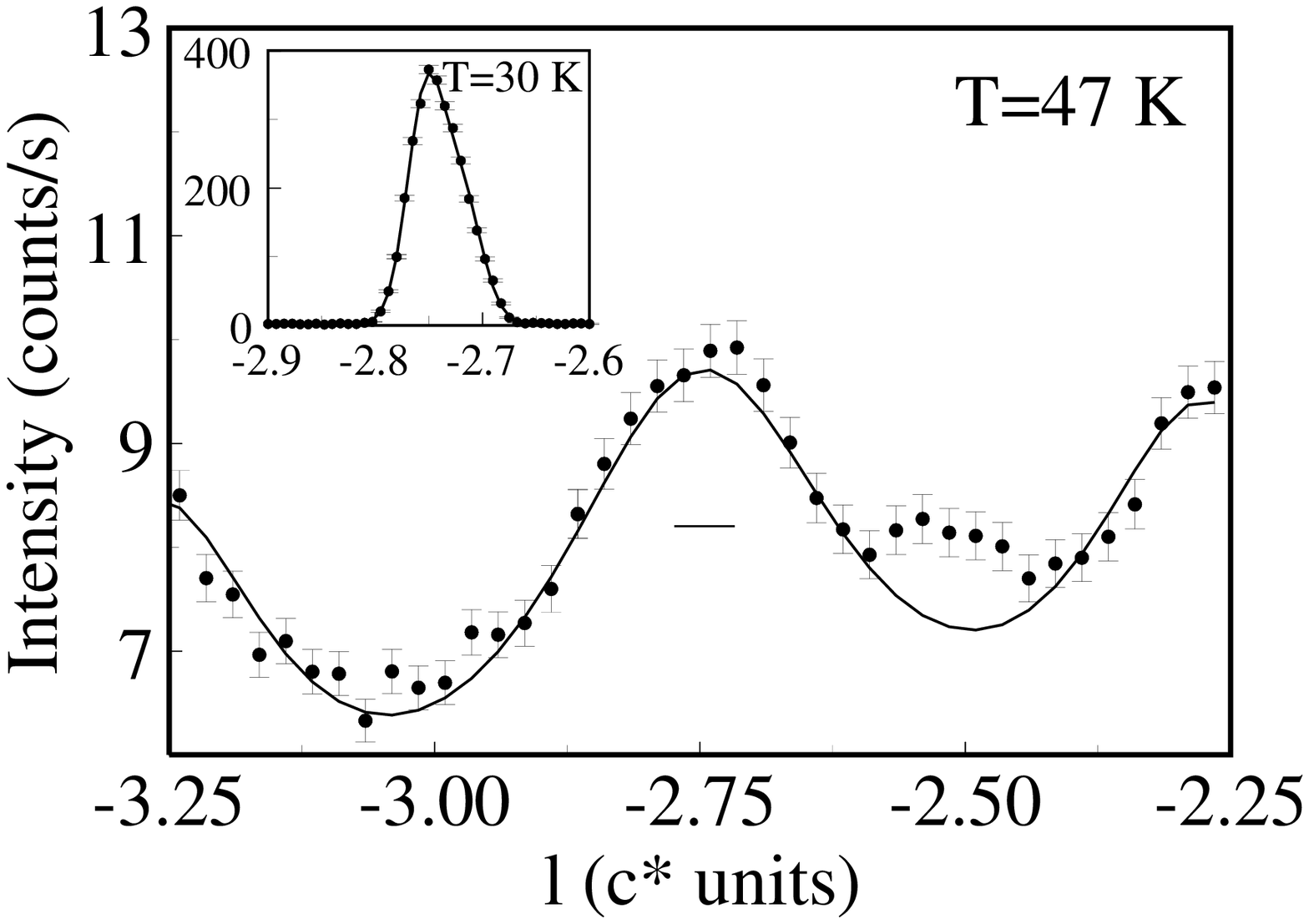,angle=0.0,height=5cm,width=7cm}}
\vskip 0.2cm
\caption{Scans of the ($-1.5,0.5,-2.75$) peak at 70 K, 90 K and 47 K in the $%
{\bf a}^{*}$ (top), ${\bf b}^{*}$ (middle) and ${\bf c}^{*}$ (bottom)
directions respectively. The inserts show the same scans at 30 K. The solid
lines are the results of the fits described in the text. The lines under the
peaks indicate the experimental resolution 2R (see text).}
\label{scans}
\end{figure}

Crystal growth was carried out by the flux method by melting under vacuum in 
platinum crucibles appropriately compacted mixtures of V$_{2}$O$_{5}$, 
V$_{2}$O$_{3}$ and NaVO$_{5}$. These melts were then slowly cooled from 1073 K
to room temperature. Depending on the cooling rate, either needle-shaped or
plated-shaped crystals were obtained. In this study, needle-shaped single
crystals were used. A preliminary x-ray photographic study with Cu K$\alpha $
radiation first confirmed the appearance of satellite reflections below $%
T_{sp}=35$ K at the reciprocal positions ${\bf (}h+\frac{1}{2}){\bf a}%
^{*}+(k+\frac{1}{2}){\bf b}^{*}+(l\pm \frac{1}{4}{\bf )c}^{*}\equiv {\bf G}%
_{hkl}\pm {\bf q}$. These new spots were found to be more intense at large
angles and low $k$. Attempts to observe diffuse scattering above the phase
transition failed however, mainly because of a large background due to the
Cu K$\alpha $-induced fluorescence of vanadium. A long crystal ($\sim $%
9x0.3x0.2 mm$^{3}$) was then attached on the cold stage of a closed-cycle He
refrigerator installed on a normal beam-lifting-detector diffractometer. The
x-ray measurements were performed using monochromatic Mo K$\alpha $
radiation. The low temperature (30 K) lattice parameters were found to be
a=11.3 \AA , b=3.6 \AA\ and c=4.75 \AA , in agreement with ref. \cite{Isobe}%
. The experimental resolution given by the Half-Width at Half-Maximum of the
intense ($-1.5,0.5,-2.75$) satellite reflection at 30 K was $R_{a}=$0.02 \AA 
$^{-1}$, $R_{b}=$0.04 \AA $^{-1}$, $R_{c}=$0.04 \AA $^{-1}$. As shown in the
inserts on figure \ref{scans}, the profiles were best fitted by a sum of two
Gaussian lines, indicating the presence of multiple components in the
crystal.

\begin{figure}
\centerline{\epsfig{file=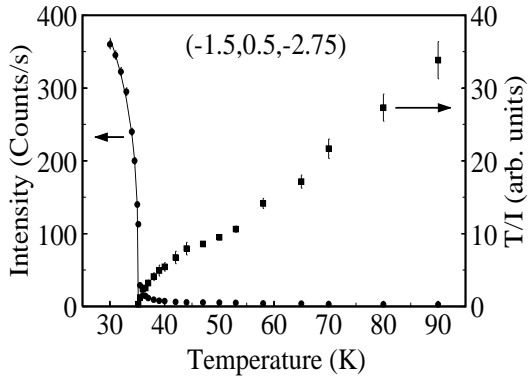,angle=0.0,height=6cm,width=7cm}}
\caption{Thermal variations of the intensity $I$ of the ($-1.5,0.5,-2.75$)
peak (right scale) and of T/I (left scale). The solid line is a fit by the
power law $(T_{sp}-T)^{0.3}$.}
\label{intens}
\end{figure}

We have measured the temperature dependence of the scattering around the
reciprocal position ($-1.5,0.5,-2.75$) from $30$ K to $90$ K by scanning in
the three directions of the reciprocal space. The peak intensity $I$ was
obtained from the ${\bf b}^{*}$-scans by subtracting the background
intensity. Figure \ref{intens} displays the temperature dependence of $I$,
proportional to the square of the structural order parameter, together with
the behavior of $T/I$, proportional to the inverse of the susceptibility
associated with this order parameter. Both quantities vanish continuously at 
$T_{sp}$, consistently with a second-order phase transition. Yet, below $%
T_{sp}$, $I(T)$ is fitted by the power law $(T-T_{sp})^{0.3}$, which leads
to a critical exponent $\beta \sim 0.15$. Although this value could correspond
to an artifact due to resolution effects close to $T_{sp}$, it is noteworthy 
that a similar fit, using the synchrotron data of Fujii $et$ $al$ \cite{Fujii} 
in the same temperature range, gives an exponent of $\beta \sim 0.2$, sizeably
different from the mean-field value 0.5 or the 0.35 exponent expected for
the behavior of an XY 3D order parameter. Interestingly, a similar value ($%
\beta =0.25$) was extracted from infrared reflectivity measurements\cite
{Smirnov}. As shown on figure \ref{intens}, the behavior of $T/I$ is also
unusual and exhibits a crossover at $\sim $50 K instead of a Curie law.

\begin{figure}
\centerline{\epsfig{file=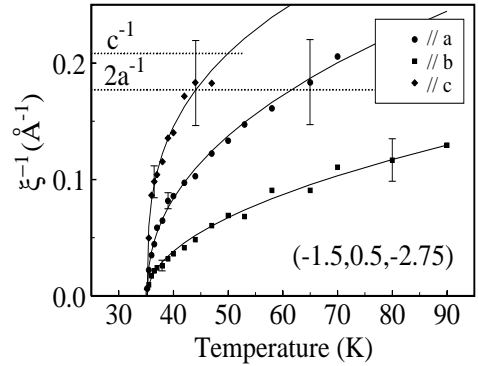,angle=0.0,height=6cm,width=7cm}}
\caption{Thermal variations of the inverse of the correlation lengths in the 
${\bf a}$, ${\bf b}$ and ${\bf c}$ directions. The solid lines are guides
for the eyes. The dashed lines indicate the inverse values of c and a/2.}
\label{correl}
\end{figure}

Above $T_{sp}$, the intrinsic correlation lengths (fig. \ref{correl}) were
obtained by a deconvolution procedure, consisting in fitting the measured
line shapes by the one-dimensional convolution of the resolution function,
obtained at low temperature, and a parametrized function that we chose to be
a Lorentzian. This procedure is known to give good results except close to $%
T_{sp}$, where the influence of the other directions becomes essential.
Figure \ref{scans} shows typical scans of the ($-1.5,0.5,-2.75$) peak at 70
K, 90 K and 47 K in the ${\bf a}^{*}$, ${\bf b}^{*}$ and ${\bf c}^{*}$
directions respectively. Due to the 1/4 value of the reduced wave vector in
the ${\bf c}^{*}$ direction, the proximity of the peaks did not allow us to 
fit the scans in this direction at temperatures larger than 47 K. A similar
problem of background subtraction occurred above 70 K in the ${\bf a}^{*}$
direction. As shown in figure \ref{scans}, a peak is still clearly visible
at 90 K ($\sim 2.5T_{sp}$) in the ${\bf b}^{*}$ direction, indicating a very
large domain of structural fluctuations. We will define a dimensional
crossover as usual when a correlation length reaches the average distance
between basic units in one direction. Within this definition, a 3D to 2D
crossover takes place at $\sim $50 K, where the extrapolated value of $\xi
_{c}$ reaches c=3.6 \AA . Because of the complex structure of $\alpha
^{\prime }$-NaV$_{2}$O$_{5}$ the actual distance between the double-rows (or
ladders) in the ${\bf a}$ direction is $a/2=5.65$ \AA . This leads us to
consider the system as quasi-1D above $\sim $60 K. It is noteworthy that $%
\xi _{b}$=7.7 \AA\ at 90 K ,\ which corresponds to twice the distance
between V atoms in the chain direction. At 40 K the anisotropy ratio ($\xi
_{b}:\xi _{a}:\xi _{c}$) is equal to (3.8 : 1.8 : 1), in qualitative
agreement with the structural anisotropy of the compound.

First of all, in the absence of a complete structure determination of the
low temperature phase, the existence of strong satellite reflections at
large angles and small $k$ qualitatively indicates the presence of a
displacive modulation with displacements mainly in the (${\bf a}$,${\bf c}$)
planes. Furthermore, it was recently concluded\cite{ILL} from a comparison
between x-ray and neutron scattering results that the vanadium atoms 
contributed the major part of this distortion. These results are consistent 
with the behavior of the high frequency dielectric constant\cite{Smirnov2}, 
from which displacements of the V atoms in the ${\bf c}$ direction have been
suggested.

Although all the thermodynamic quantities extracted from x-ray scattering
data (namely the order parameter, the inverse of its associated susceptibility
and the correlation lengths) continuously vanish at $T_{sp}$, the occurrence
of a slightly first-order transition is not ruled out\cite{Köppen}. Indeed,
the measurement of a small $\beta $ value is known to account for the small
jump of the order parameter at the phase transition. In this respect, a
strong coupling to lattice strains in the ${\bf a}$ and ${\bf c}$ directions
has been measured\cite{Köppen}, which, according to simple Landau theory can
lead to a first-order phase transition. Precise synchrotron measurements will
be performed to confirm this point. More puzzling is the crossover at $%
\sim $50 K in the behavior of $T/I$. A similar anomaly was already observed
in MEM(TCNQ)$_{2}$\cite{Liu2}, where isotropic lattice fluctuations exist
from $T_{c}=$18 K to 300 K, but was interpreted as due to the appearance of
critical SP fluctuations locking in on a 'preexistent' soft mode, so far
unexplained. Indeed, lattice fluctuations were observed even for T$%
\gtrsim J$, {\it i.e.} above the regime of quantum antiferromagnetic
fluctuations. This is not the case for $\alpha ^{\prime }$-NaV$_{2}$O$_{5}$,
where the structural fluctuations are measured for T$\lesssim J=$280 K. The
anomaly of the thermal dependence of $T/I$ is rather to be compared with the
one observed in CuGeO$_{3}$\cite{Schoeffel}. As in this compound, it appears
at the 3D-2D crossover, that we have also observed at $\sim $50 K.

The anisotropy ratio $\xi _{b}:\xi _{a}:\xi _{c}$ = 3.8 : 1.8 : 1 is similar
to the one observed in the Peierls transition of the blue bronze K$_{0.3}$MoO%
$_{3}$ (7.5 : 1.8 : 1)\cite{Girault}, or the SP transition of (BCP-TTF)$_{2}$%
AsF$_{6}$\cite{Liu2}( 3.6 : 2.6 : 1). This slight anisotropy is also to be
compared with the larger one observed in the magnetic excitation spectrum by
neutron scattering experiments\cite{Yosihama} at 40 K. This indicates
firstly that transverse interactions other than magnetic must be considered
in order to understand the structural fluctuations. Moreover, the stabilization 
of the transverse components $q_{{\bf a}*}=1/2$ and $q_{{\bf c}*}=1/4$ requires
next-nearest neighbor interactions between ladders. The long-range character
of Coulomb interactions involved in the charge ordering process gives here a
natural way to introduce such interactions. As far as the lattice is
concerned, the coupling with the charge ordering is expected to occur
through the size difference between V$^{4+}$ and V$^{5+}$\cite{Mostovoy}.
Nevertheless, the exact charge configuration is difficult to calculate and
it has been suggested that the 'zig-zag' ordering could be stabilized {\it %
via} the lattice distortion\cite{Mostovoy}. The role of the spin gap in the
stabilization of the low temperature phase could also be essential. It has
been shown from symmetry reasons that in the zig-zag ordering, the existence
of alternate exchange interactions responsible for the observed spin gap
cannot be realized if $q_{{\bf a}*}=0$ but only if $q_{{\bf a}*}=1/2$,
thus coupling {\it via} second neighbors the charge ordering and the spin 
gap\cite {Mostovoy}. Seo {\it et al}\cite{Seo} rather considered an 
interladder pairing for this order, {\it i.e. via} first neighbors. In both 
cases, the spin gap is a consequence of the transverse ordering. According 
to these theories, the local pairing above $T_{sp}$ would become effective 
when the {\it transverse} correlation length $\xi _{a}\ $exceeds $a$ or $a/2$
(depending on the model considered), {\it i.e.} for $T\lesssim $40 K or 60
K. In this respect, deviation from the BF law should be visible in the 40-60
K temperature range, which can be guessed in the data published so far\cite
{Isobe} but requires further investigations. It might also explain the
anomalous behavior of T/I at $\sim $50 K. This is in strong contrast with
the conventional SP scenario, in which the pairing becomes effective when
the correlation lengths exceed the distance between spins {\it along} the
chain\cite{Dumoulin}. From our measurements, this regime would occur below a
temperature much higher than 90 K and a deviation to the BF law should have
been detected. Besides, it is worth pointing out that precursor effects have
been observed on the dielectric constant temperature dependence at least 8K 
above the phase transition\cite{Smirnov2} and on the $C_{22}$ elastic 
constant variations 70 K above $ T_{sp}$\cite{Fertey}.

As far as the BCS-ratio is concerned, it is instructive to compare the
values $\sim $7\cite{JPP} and $\sim $6.5 measured in the blue bronze and $%
\alpha ^{\prime }$-NaV$_{2}$O$_{5}$, respectively. In the blue bronze this
deviation from the mean-field value is interpreted as being due to the 
existence of a strong regime of quasi-1D structural fluctuation, depressing 
the phase transition temperature\cite{JPP}. The same argument can be used in 
our case. This makes the BCS-ratio value less anomalous but raises again the 
question of the relation between spin gap and charge ordering, and 
of the nature of the transverse interactions we have shown to be essential 
in this new type of magneto-elastic phase transition.

S. R. acknowledges the help of S. Besse for the experiments and useful
discussions with P. Lederer, R. Moret and J.-P Pouget.


\begin{references}
\bibitem{Hui}  S. Huizinga {\it et al.}, Phys. Rev. B{\bf \ 19}, 4723 (1979).

\bibitem{Bray}  J. M. Bray {\it et al.}, in {\it Extended Linear Chain
Compounds}, edited by J.S. Miller (Plenum, New York, 1983) Vol. 3, p. 353.

\bibitem{Monc}  D. E. Moncton {\it et al.}, Phys. Rev. Lett. {\bf 39}, 507
(1977).

\bibitem{Creuzet}  F. Creuzet {\it et al.}, Synth. Met. {\bf 19}, 289
(1987); J.-P. Pouget {\it et al.}, Mol. Crys. Liq. Cryst. {\bf 79}, 129
(1982).

\bibitem{Laver}  R. Laversanne {\it et al.}, Mol. Crys. Liq. Crys. {\bf 119}%
, 317 (1985).

\bibitem{Ducasse}  L. Ducasse {\it et al.}, Synth. Met. {\bf 27}, B543
(1988).

\bibitem{Liu}  Q. Liu {\it et al.}, Synth. Met. {\bf 42}, 1879 (1991).

\bibitem{Hase}  M. Hase, I. Terasaki and K. Uchinokura, Phys. Rev. Lett. 
{\bf 70}, 3651 (1993) and for a review J.P. Boucher and L.P. Regnault, J.
Phys. I France {\bf 6}, 1939 (1996).

\bibitem{Liu2}  Q. Liu {\it et al.}, Synth. Met. {\bf 56}, 1840 (1993).

\bibitem{Dumoulin}  B. Dumoulin {\it et al.}, Phys. Rev. Lett {\bf 76}, 1360
(1996).

\bibitem{Schoeffel}  J.-P. Schoeffel, J.-P. Pouget, G. Dhalenne and A.
Revcolevschi, Phys. Rev. B {\bf 53}, 14971 (1996)

\bibitem{Isobe}  M. Isobe and Y. Ueda, J. Phys. Soc. Japan {\bf 65,} 1178
(1996).

\bibitem{Fujii}  Y. Fujii {\it et al.}, J. Phys. Soc. Japan {\bf 66,} 326
(1997).

\bibitem{Carpy}  P. A. Carpy and J. Galy, Acta Crist. Sect. B {\bf 31,} 1481
(1975).

\bibitem{Centro}  A. Meetsma {\it et al.}, Acta Cryst., in press; H. G. von
Schnering {\it et al}, Z. Kristallogr. {\bf 213}, 246 (1998).

\bibitem{Smolinski}  H. Smolinski {\it et al.}, cond-mat/9801276.

\bibitem{Omaha}  T. Omaha, H. Yasuoka, M. Isobe and Y. Ueda, preprint

\bibitem{Köppen}  M. K\"{o}ppen {\it et al.}, Phys. Rev. B {\bf 57}, 8466
(1998).

\bibitem{Seo}  H. Seo and H. Fukuyama, cond-mat/9805185.

\bibitem{Thalmeier}  P. Thalmeier and P. Fulde, cond-mat/9805230.

\bibitem{Nishimoto}  S. Nishimoto and Y. Ohta, cond-mat/9805336.

\bibitem{Mostovoy}  M. V. Mostovoy and D. I. Khomskii, cond-mat/9806215.

\bibitem{Smirnov}  D. Smirnov {\it et al.}, Phys. Rev. B {\bf 57}, 11035
(1998).

\bibitem{ILL}  T. Chatterji {\it et al.}, Solid State Commun. {\bf 108}, 23
(1998)

\bibitem{Smirnov2}  A. I. Smirnov {\it et al.}, cond-mat/9808165.

\bibitem{Yosihama}  T. Yoshihama {\it et al.}, J. Phys. Soc. Jpn {\bf 67},
744 (1998).

\bibitem{Girault}  S. Girault, A.H. Moudden and J.-P. Pouget, Phys. Rev. B%
{\bf \ 39}, 4430 (1989).

\bibitem{JPP}  For a discussion see {\it e.g. }J.-P. Pouget, in {\it %
Low-Dimensionnal Electronic properties of Molybdenum Bronzes and Oxides},
edited by C. Schlenker (Kluwer Academic Publishers, 1989) p. 87.

\bibitem{Fertey}  P. Fertey {\it et al.}, Phys. Rev. B {\bf 57, }13698
(1998).
\end{references}
\end{document}